**Title:** Damage to white matter bottlenecks contributes to language impairments after left hemispheric stroke

**Authors** Joseph C. Griffis[1], Rodolphe Nenert[2], Jane B. Allendorfer[2], Jerzy P. Szaflarski[2]

**Institutional Affiliations:** University of Alabama at Birmingham Department of Psychology[1], University of Alabama at Birmingham Department of Neurology[2]

**Corresponding Author Information:** Joseph C. Griffis (joegriff@uab.edu)
Department of Neurology and UABEC, University of Alabama at Birmingham, 312 Civitan International Research Center, 1719, 6th Avenue South, Birmingham, AL 35294-0021

**Abstract**

Lesion-behavior studies indicate that damage to the white matter underlying the left posterior temporal lobe leads to deficits in multiple language functions. Based on evidence that multiple long-range white matter tracts pass through this region, it has been proposed that the posterior temporal white matter may correspond to a bottleneck where multiple long-range projections associated with both dorsal and ventral language pathways are vulnerable to simultaneous damage. Damage to a second putative white matter bottleneck in the left deep prefrontal white matter involving projections associated with ventral language pathways and thalamo-cortical projections has recently been proposed as a source of semantic deficits after stroke. However, the effects of damage to *a priori* identified white matter bottlenecks on language function have not been directly investigated. Here, we first used white matter atlases to identify the previously described white matter bottlenecks in the posterior temporal and deep prefrontal white matter. We then assessed the effects of damage to each bottleneck region on measures of category fluency, picture naming, and auditory semantic decision-making in 43 left hemispheric stroke patients with chronic aphasia. Damage to the posterior temporal bottleneck predicted deficits on all language measures, while damage to the anterior bottleneck only predicted deficits in category fluency. Importantly, the effects of damage to the bottleneck regions were not attributable to lesion volume, lesion loads on the tracts traversing the bottlenecks, or damage to nearby cortical language areas. Multivariate lesion-symptom mapping and fiber tracking analyses corroborated these findings, and identified additional cortical and white matter lesion predictors of chronic language deficits. Together, our results provide strong support for the proposal that spatially specific white matter damage affecting white matter bottlenecks, particularly in the posterior temporal lobe, may represent an often overlooked source of chronic language deficits in patients with left hemisphere stroke. Our results suggest that damage to this area is likely to simultaneously disrupt signaling via the simultaneous disruption of dorsal and ventral language processing streams. Our findings highlight the importance of utilizing spatially sensitive measures of white matter damage in lesion-behavior research.

**Keywords:** aphasia, stroke, white matter, language, MRI

**Abbreviations:** angular gyrus (AG), anterior temporal lobe (ATL), anterior thalamic radiations (ATR), arcuate faciculus (AF), auditory semantic decisions (AudSem), Boston Naming Test (BNT), Controlled Oral Word Association Test (COWAT), direct total lesion volume control (DTLVC), false discovery rate (FDR), family-wise error rate (FWE), frontal aslant tract (FAS), grey matter (GM), inferior frontal gyrus pars opercularis (IFGpop), inferior frontal gyrus pars triangualris (IFGptr), inferior fronto-occipital fasciculus (IFOF), inferior longitudinal fasciculus (ILF), middle longitudinal fasciculus (MDLF), posterior middle temporal gyrus (pMTG), posterior superior temporal gyrus (pSTG), posterior supramarginal gyrus (pSMG), precentral gyrus (PCG), region of interest (ROI), Semantic Fluency Task (SFT), support vector regression lesion-symptom mapping (SVR-LSM), temporo-occipital middle temporal gyrus (tocMTG), tissue probability map (TPM), white matter (WM)

# 1. Introduction

Damage to long-range WM pathways likely contributes substantially to language deficits after left hemisphere stroke (Catani and Mesulam, 2008). Damage to the WM underlying the left pSTG/pMTG is consistently implicated as a source of deficits in multiple language domains including comprehension (Dronkers *et al.*, 2004; Geva *et al.*, 2012; Henseler *et al.*, 2014), naming (Baldo *et al.*, 2013; Harvey and Schnur, 2015), repetition (Butler *et al.*, 2014; Henseler *et al.*, 2014), and phonology (Butler *et al.*, 2014). Lesions disrupting middle temporal connections also predict poor responses to language therapies (Bonilha *et al.*, 2015; Fridriksson, 2010).

Why might damage to the WM in this area have such broadly negative impacts on language outcomes? Portions of the WM under the pSTG/pMTG contain projections associated with multiple tracts (Turken and Dronkers, 2011), including dorsal (sensori-motor) and ventral (associative) language pathways (Kümmerer *et al.*, 2013). Fibers associated with at least three language-relevant tracts traverse this region – the AF (dorsal stream), ILF (ventral stream), and IFOF (ventral stream) (Catani and Mesulam, 2008; Turken and Dronkers, 2011). Thus, it has been proposed that this corresponds to a structural weak point, or "bottleneck", where multiple pathways are vulnerable to simultaneous disruption by focal damage (Turken and Dronkers, 2011). The observation that fibers associated with the ATT (thalamo-cortical), UF (ventral stream), and IFOF (ventral stream) form a bottleneck in the prefrontal WM near areas where damage is associated with chronic deficits in semantic recognition supports the proposal that damage to bottleneck regions may play a role in chronic language deficits after stroke (Mirman, *et al.*, 2015a).

However, the conclusions that can be drawn from these previous studies are limited because both bottleneck regions were identified as part of *post-hoc* exploratory analyses based on the results of voxel-wise lesion-deficit mapping (Dronkers *et al.*, 2004; Mirman *et al.*, 2015a). The effects of damage to *a priori* identified bottlenecks in these regions on language outcomes have not been investigated. We aimed to bridge this gap by characterizing how deficits in measures of category fluency, picture naming, and auditory semantic comprehension relate to lesions affecting the bottleneck regions described by previous reports (Mirman et al., 2015a; Turken and Dronkers, 2011). We expected that damage to the bottleneck underlying the left pSTG/pMTG would be associated with chronic impairments on all language measures, as broad deficits would be expected to follow the simultaneous disruption of both ventral and dorsal language streams. Based on the report by Mirman and colleagues (2015a), we expected that damage to the prefrontal bottleneck would be associated with deficits in picture naming and category fluency. To enable strong conclusions about our specific findings, we demonstrate that the effects of damage to these regions are not attributable to lesion loads on the tracts traversing them or to concomitant cortical damage. To assess the robustness of our results to a data-driven approach and thoroughly characterize lesion-deficit relationships, we utilized SVR-LSM with DTLVC (Zhang *et al*., 2014). Exploratory fiber tracking enabled further characterization of lesion effects on inter-regional connections.

## 2. Methods

*2.1 Participants*

All procedures were approved by the Institutional Review Boards of the participating institutions and performed in accordance with Declaration of Helsinki ethics principles and principles of informed consent. Data were prospectively collected from 43 patients with chronic (> 1 year) post-stroke aphasia participating in different studies by our laboratory. Patients were excluded if they had diagnoses of degenerative/metabolic disorders, diagnoses of severe depression or other psychiatric disorders, were pregnant, were not fluent in English, or had any contraindication to MRI/fMRI. All patients had a single left hemispheric stroke resulting in aphasia at least 1 year prior to data collection. No patients had right hemispheric damage. The mean patient age was 53 (SD=15), 25 patients were male, and the mean pre-stroke handedness as determined by the Edinburgh Handedness Inventory (Oldfield, 1971) was 0.85 (SD=0.43). A detailed characterization of patient characteristics is provided in Supplementary Table 1.

*2.2 Neuroimaging data collection*

MRI Data were acquired at the University of Alabama at Birmingham using a 3T head-only Siemens Magnetom Allegra scanner provided by the Civitan International Research Center Functional Imaging Laboratory. These data consisted of 3D high-resolution T1-weighted anatomical scans (TR/TE = 2.3 s/2.17 ms, FOV = 25.6×25.6×19.2 cm, matrix = 256x256, flip angle = 9 degrees, slice thickness = 1mm). MRI data were also collected at the Cincinnati Children's Hospital Medical Center on a 3T research-dedicated Philips MRI system provided by the Imaging Research Center. These data consisted of 3D high-resolution T1-weighted anatomical scans (TR/TE = 8.1 s/2.17 ms, FOV = 25.0×21.0×18.0 cm, matrix = 252x211, flip angle = 8 degrees, slice thickness = 1mm).

*2.3 Lesion identification*

MRI data were processed using Statistical Parametric Mapping (SPM) (Friston *et al.*, 1995) version 12 running in MATLAB r2014b (The MathWorks, Natick MA, USA). T1-weighted scans were segmented and normalized using the unified normalization procedure implemented in SPM12. Lesion probability maps were created for each patient using a probabilistic lesion classification algorithm implemented in the *lesion_gnb* toolbox for SPM12 (Griffis *et al.*, 2016a). Lesion probability maps were manually thresholded to ensure that the resulting binary masks precisely reflected the lesions, and then resampled to 2 mm isotropic resolution. Lesion frequencies are shown in Figure 1A.

*2.4 Language measures*

Patients completed a battery of language assessments that included the BNT (Kaplan *et al.*, 2001), SFT form A (Kozora and Cullum, 1995), and COWAT form A (Lezak *et al.*, 1995) prior to MRI scanning. The BNT is a picture naming test that consists of black and white line drawings of animate and inanimate items. Naming abilities were measured as the number of correctly named pictures. The SFT and COWAT both involve generating words in response to a given prompt within a one minute time limit. The SFT is a measure of semantic fluency that uses semantic category prompts (animals, fruits/vegetables, things that are hot), whereas the COWAT is a measure of phonological fluency that uses letter category prompts (C, F, and L).

The AudSem task was completed in the scanner, and this was used as a measure of auditory semantic comprehension abilities. This task robustly activates networks involved in semantic comprehension (Binder *et al.*, 1997) even in patients with chronic stroke (Eaton *et al.*, 2008). The fMRI data associated with this task are characterized separately (Griffis et al., 2016b), as this study is focused on structural data only. Participants heard 80 spoken English nouns designating different animals, and decided if the animals were both "native to the United States" and "commonly used by humans" (e.g. for food, clothing, or labor), indicated by button press. A mock run with five trials outside of the scanner confirmed that patients understood the task. AudSem task data were missing for 4 patients due to hardware issues.

Scores on the SFT and COWAT were strongly correlated ($r=0.92$), and patients' general verbal fluency was measured as the average number of words generated for both the SFT and COWAT. A similar composite fluency outcome was used by a recent study investigating the effects of tract disconnection on language abilities in chronic stroke patients (Hope *et al.*, 2015). Scores on the BNT ($r=0.76$) and semantic decision task ($r=0.53$) showed positive correlations to scores on the composite fluency measure. Scores on the BNT were also positively correlated with scores on the AudSem task ($r=0.43$). Language scores are shown in Figure 1B. Supplemental analyses considering the COWAT and SFT separately are provided in Supplementary Analysis 1.

2.5 White matter bottleneck ROIs

We defined two *a priori* ROIs corresponding to the putative bottleneck regions described by previous studies (Mirman, *et al.*, 2015a,b; Turken and Dronkers, 2011). We first defined an anterior bottleneck ROI according to precisely the same procedure used by Mirman and colleagues to identify the bottleneck region described in their reports (2015a; 2015b). This procedure consisted of thresholding probabilistic atlas labels (ICBM-DTI atlas – available at http://fsl.fmrib.ox.ac.uk/fsl/fslwiki/Atlases (Oishi *et al.*, 2008)) for the left UF, IFOF, and ATR at a 20% tract probability threshold, and intersecting the thresholded labels. Voxels contained within the intersection of these tract labels were defined as the anterior bottleneck ROI. Next, we defined a posterior bottleneck ROI according to the same procedure, but using labels for the three previously identified in the posterior temporal WM language-relevant tracts: the left ILF, IFOF, and SLF (Turken and Dronkers, 2011). To emphasize, these tracts were chosen based on reports that fibers associated with these tracts traverse portions of the WM underlying the left pMTG/pSTG where lesions are associated with impaired comprehension (Turken and Dronkers, 2011; Dronkers et al., 2004). While additional tracts (i.e. MDLF and tapetum) were also reported to traverse the white matter near this region (Turken and Dronkers, 2011), these regions were not included in the ROI definition because probabilistic labels for these tracts were not available, and their importance for language is not clear. Both ROIs were resampled to 2 mm isotropic resolution. After resampling, the posterior ROI contained 47 voxels, and the anterior ROI contained 66 voxels. The process used to define each ROI is outlined in Figure 2A.

*2.6 Multiple regression analyses*

Multiple linear regressions assessed whether damage to each bottleneck ROI predicted chronic language deficits. Binary lesion status (lesion vs. no lesion) was used to designate lesion effects at each ROI (rather than a continuous measure such as percent lesion overlap) because each ROI contained only a small number of voxels. 14 patients did not have damage at either ROI, 10 patients had damage to only the anterior ROI, 4 patients had damage to only the posterior ROI, and 15 patients had damage to both ROIs. Representative slices with lesion frequencies for each subgroup of patients are shown in Figure 2B. As noted in Section 2.4, 4 patients were excluded from the AudSem analyses due to hardware issues that resulted in a failure to collect behavioral data in-scanner; 1 of these patients did not have damage at either ROI, 1 had damage to both ROIs, and 2 had damage to only the anterior ROI.

Analyses proceeded by first fitting a bottleneck predictor model for each outcome. If significant bottleneck predictors were identified in this model, then a second model was fit using hierarchical regression. The bottleneck predictor model determined if damage to the bottleneck ROIs significantly predicted the language outcome while controlling for lesion volume. The hierarchical model was fit in two stages. The first (control model) accounted for (1) the effects of damage to the tracts used to define each significant bottleneck ROI predictor from the first model, and (2) the effects of damage to cortical language areas near each significant bottleneck ROI predictor from the first model. The second (control+bottleneck model) tested whether the addition of significant bottleneck ROI predictors from the bottleneck predictor model significantly improved the fit of the control model model. This approach is conceptually similar to the approach used by Hope and colleagues (2015) in their comparison of tract lesion load and tract disconnection measures as predictors of chronic language outcomes. Our approach is also more comprehensive than most studies investigating WM lesion contributions to chronic language deficits, as controls for concomitant cortical damage are not typically employed (Forkel *et al*., 2014; Geva *et al*., 2015; Hope *et al*., 2015; Ivanova *et al*., 2016; Kümmerer *et al*., 2013; Marchina *et al*., 2011). To assess whether the conclusions drawn from these analyses would be robust against changes in ROI definition, control analyses were performed using ROIs defined using an alternate white matter atlas (Rojkova *et al*., 2016), and are provided as supplementary material.

For the bottleneck predictor model of each language outcome, the standardized language outcome was entered as the dependent variable, and 4 predictor variables were simultaneously entered: (1) standardized lesion volume, (2) lesion status at the anterior bottleneck ROI, (3) lesion status at the posterior bottleneck ROI, and (4) the lesion status interaction term. These models were considered significant if the *p*-value for the *F*-test against the intercept-only model survived Bonferroni-Holm correction to control the FWE at 0.05 across all three models (Aickin and Gensler, 1996). Parameter estimates for the variables in model 1 were considered significant if the *p*-value of the test statistic survived Bonferroni-Holm correction to control the FWE at 0.05 across all 4 variables in the model. The control and control+bottleneck models were fit in a hierarchical multiple regression if significant bottleneck ROI predictors were identified in model 1. To enable these control analyses, additional predictors were defined as described in the following paragraphs.

Tract and cortical lesion loads were calculated for use as predictors in the hierarchical regressions. Tract control ROIs were defined as the 20% thresholded atlas labels used to define the ROIs. Cortical control ROIs were defined using the Harvard-Oxford maximum probability cortical atlas to control for damage to language-relevant areas with close proximity to each bottleneck ROI. Cortical ROIs to control for the anterior bottleneck ROI were defined as the left IFGptr, IFGpop, PCG, and insula. Cortical ROIs to control for the posterior bottleneck ROI were defined as the left pSTG, pMTG, tocMTG, pSMG, and AG. These regions were chosen based on their close proximity to each ROI, and because they are generally implicated in language processing (e.g. Fedorenko & Thompson-Schill, 2014; Friederici & Gierhan, 2013; Price, 2010). Cortical ROIs were masked to include only voxels with a grey matter tissue probability of at least 20% as determined using the GM TPM included in SPM12, and are shown in Figure 2C. Lesion loads were calculated as the proportion of voxels in each control ROI overlapped with each patient's binary lesion mask (excluding voxels corresponding to the bottleneck ROIs).

The control model was fit in the first block of the hierarchical regression by entering lesion load predictors for (1) the tracts used to define significant bottleneck ROI predictor(s) from the bottleneck model, and (2) the cortical control ROIs for significant bottleneck ROI predictor(s) the bottleneck model. For the control+bottleneck model, the significant bottleneck ROI predictor(s) from the bottleneck model were added in a second block to the control model, and the change in model $R^2$ was evaluated to determine whether the addition of the bottleneck ROI predictor(s) significantly improved model fit. Changes in model $R^2$ were considered significant if the $p$-value for the $F$-test on the $R^2$ change statistic survived Bonferronni-Holm correction to control the FWE at 0.05 across all three $R^2$ change tests (one for each language measure).

2.7 Multivariate lesion-symptom mapping with lesion volume control

Multivariate lesion-symptom mapping was performed using the SVR-LSM toolbox for SPM (Zhang *et al*., 2014). Unlike traditional lesion-symptom mapping approaches that separately test the effects of damage to each voxel on behavioral scores, SVR-LSM identifies lesion-behavior relationships at all voxels simultaneously, making it more sensitive for detecting lesion-symptom relationships than traditional approaches (Zhang *et al*., 2014). To control for lesion volume effects, we utilized the DTLVC option in the SVR-LSM toolbox. This method normalizes each patient's lesion map to have a unit norm, such that the lesion status of each voxel is equal to either 0 or the reciprocal of the norm ($1/\sqrt{\text{lesion volume}}$), and provides better control than the regression of behavioral scores on lesion volume (Zhang *et al*., 2014). Analyses without lesion volume control are provided as supplements (Supplementary Material 2). SVR-LSM+DTLVC analyses only considered voxels that were lesioned in at least 10 patients. While there is not a consensus regarding whether SVR-LSM requires multiple comparisons correction (Mirman *et al.,* 2015b), lesion-behavior relationships that survived a False Discovery Rate (Benjamini and Hochberg, 1995; Genovese *et al.*, 2002) (FDR) correction threshold of 0.10 as determined by 2000 permutation tests are reported here (Mirman *et al.*, 2015b). We note that unlike FWE-controlling procedures, FDR correction estimates the proportion of discoveries that are likely to be false, and an FDR threshold of 0.10 indicates that on average no more than 10% of significant voxels are false discoveries (Benjamini and Hochberg, 1995). Additionally, unlike predictive SVR, where a model is trained to predict outcomes for new cases, the SVR-LSM model is intended to identify significant lesion predictors of behavioral outcomes and is fit to the entire dataset (Zhang *et al.*, 2014). We used empirically optimized values for the model parameters C (C=30) and gamma (gamma=5) as reported by Zhang and colleagues (2014) in their validation and optimization of the SVR-LSM method.

*2.8 Exploratory deterministic tractography*

Tractography was performed to characterize the pathways affected by WM lesions found to most strongly predict chronic language impairments in the SVR-LSM+DTLVC analyses. The FDR-thresholded maps were masked to include only voxels with a WM tissue probability of at least 20% as determined by the WM TPM included in SPM12, and up to three ROIs per language outcome were defined using the default ROI creation (12mm spherical ROIs masked to exclude non-significant voxels) and peak reporting (top 3 peaks with minimum distances of 30mm) settings in the bspmview toolbox for SPM12 (http://www.bobspunt.com/bspmview/). This resulted in up to three ROIs corresponding to the top WM predictors for each language outcome.

Deterministic fiber tracking utilized a publicly available group-averaged tractography atlas (WU-Minn HCP Consortium; HCP-842 atlas - http://dsi-studio.labsolver.org/download-images/hcp-842-template) from 842 healthy individuals' diffusion MRI data from the Human Connectome Project (2015 Q4, 900 subject release). Data were accessed under the WU-Minn HCP open access data use term. Data acquisition utilized a multi-shell diffusion scheme (b-values: 1000, 2000, and 3000 s/mm2; diffusion sampling directions: 90, 90, and 90; in-plane resolution: 1.25mm). Data were reconstructed in MNI template space using Q-space diffeomorphic reconstruction (QSDR) as implemented in DSI_Studio (Yeh and Tseng, 2011) to obtain the spin distribution function (Yeh *et al.*, 2010) (diffusion length sampling ratio: 1.25; output resolution: 2mm). Deterministic fiber tracking (Yeh *et al.*, 2013) proceeded by seeding the whole brain to calculate 100,000 tracts terminating within the 20% thresholded GM TPM included with SPM12, and used default tracking parameters implemented in DSI_studio (angular threshold: 60 degrees; step size: 1 mm; quantitative anisotropy threshold determined automatically by DSI Studio to be 0.24; tracks with length less than 30 mm were discarded). The resulting tracts were filtered to leave only tracts that passed through each ROI. The filtered tracts were manually separated and labeled according to previous reports (Catani and Mesulam, 2008; Catani and Thiebaut de Schotten, 2008; Catani *et al.*, 2002, 2013; Hua *et al.*, 2008; Turken and Dronkers, 2011).

**3. Results**

*3.1 Multiple regression analyses*

*3.1.1 Fluency results*

Inspection of fitted residual and normal probability plots for multiple regression for the fluency measure indicated violations of homoscedasticity. To correct for this, a constant of 1 was added to fluency scores and the resulting values were transformed using a square root transformation; inspection of fitted residuals and normal probability plots indicated that this transformation successfully corrected the violated assumptions. Notably, the untransformed and transformed fluency measures were highly correlated ($r=0.97$), so the results of the analyses using the transformed variable can be interpreted in a straightforward way.

The bottleneck predictor model significantly predicted fluency scores ($R^2=0.59$, $F_{4,38}=13.7$, $p<0.001$, corrected). Similar results were obtained when the model was fit to the untransformed scores ($R^2=0.58$, $F_{4,38}=13.1$, $p<0.001$, corrected). Anterior lesion status, posterior lesion status, and the interaction term each uniquely predicted fluency scores (see Figure 3 A-B). Because significant effects were revealed for the anterior bottleneck ROI, posterior bottleneck ROI, and interaction term, we next fit the control model. The control model included lesion loads on the SLF/AF, ILF, IFOF, UF, and ATR as predictors to account for the effects of concomitant damage to the tracts used to define each bottleneck ROI, and lesion loads on both sets of cortical control ROIs as predictors to account for the effects of concomitant cortical damage to language-relevant areas proximal to each bottleneck ROI (Figure 3A). The control model predicted fluency scores ($R^2$=0.64, $F_{14,28}$=3.52, $p$=0.002). Next, to determine if the significant bottleneck ROI predictors from the bottleneck model provided additional information about fluency scores, anterior ROI lesion status, posterior ROI lesion status, and the interaction term were added to the control model, resulting in the control+bottleneck model (Figure 3A). The control+bottleneck model explained significantly more variance than the control model ($R^2$=0.84, $\Delta R^2$=0.20, $F_{3,25}$=10.42, $p$=0.0001, corrected), indicating that ROI lesion status provided unique information about fluency scores even after accounting for concomitant cortical and tract damage. Plots illustrating the difference in model fit between the control and control+bottleneck models are shown in Figure 3C.

*3.1.2 Naming results*

The bottleneck predictor model significantly predicted naming scores ($R^2$=0.51, $F_{4,38}$=9.95, $p$<0.001, corrected). Only posterior lesion status uniquely predicted fluency scores (see Figure 3 A-B). Because significant effects were revealed for the posterior bottleneck ROI in the bottleneck predictor model, we next fit the control model. The control model included lesion loads on the SLF/AF, ILF, and IFOF as predictors to account for the effects of concomitant damage to the tracts used to define the posterior bottleneck ROI, and lesion loads the posterior cortical control ROIs as predictors to account for the effects of concomitant cortical damage to language-relevant areas proximal to the posterior bottleneck ROI (Figure 3A). The control model predicted naming scores ($R^2$=0.49, $F_{8,34}$=4.04, $p$=0.0018). Next, to determine if the significant bottleneck ROI predictor from the bottleneck predictor model provided additional information about naming scores, posterior ROI lesion status was added to the control model, resulting in the control+bottleneck model (Figure 3A). The control+bottleneck model explained significantly more variance than the control model ($R^2$=0.72, $\Delta R^2$= 0.23, $F_{1,33}$=28.03, $p$<0.001, corrected), indicating that ROI lesion status provided unique information about naming scores even after accounting for concomitant cortical and tract damage. Plots illustrating the difference in model fit between the control and control+bottleneck models are shown in Figure 3C.

*3.1.3 AudSem results*

The bottleneck predictor model significantly predicted AudSem scores ($R^2$=0.41, $F_{4,34}$=5.91, $p$=0.001, corrected). Only posterior lesion status uniquely predicted AudSem scores (Figure 3A-B). Because significant effects were revealed for the posterior bottleneck ROI in the bottleneck predictor model, we next fit the control model. The control model included lesion loads on the SLF/AF, ILF, and IFOF as predictors to account for the effects of concomitant damage to the tracts used to define the posterior bottleneck ROI, and lesion loads the posterior cortical control ROIs as predictors to account for the effects of concomitant cortical damage to language-relevant areas proximal to the posterior bottleneck ROI (Figure 3A). The control model did not strongly predict AudSem scores ($R^2$=0.34, $F_{8,30}$=1.91, $p$=0.09). Next, to determine whether the significant bottleneck ROI predictor from the bottleneck predictor model provided additional information about AudSem scores, posterior ROI lesion status was added to the control model, resulting in the control+bottleneck model (Figure 3A). The control+bottleneck model explained significantly more variance than the control model ($R^2$=0.50, $\Delta R^2$ change=0.16, $F_{1,29}$=9.28, $p$=0.005, corrected), indicating that ROI lesion status provided unique information about AudSem scores even after accounting for concomitant cortical and tract damage. Plots illustrating the difference in model fit between the control and control+bottleneck models are shown in Figure 3C.

*3.2 Multivariate lesion symptom mapping with lesion volume control*

Significant lesion predictors of deficits in each of the language measures are shown in Figure 4A, and cluster peaks are provided in the Table. Figure 4B illustrates the overlap of the SVR-LSM FDR maps with each of the bottleneck ROIs. Note that while both ROIs overlapped with the map for fluency, only the posterior bottleneck ROI overlapped with the maps for naming and auditory semantic decisions, corroborating the results of our *a priori* analyses.

Three WM peaks meeting our criteria were identified for fluency and naming, and one WM peak was identified for AudSem. Tractography of the derived ROIs is shown in Figure 5B.

*3.3 Deterministic tractography*

As described in the methods section, ROIs were created on the white matter peak regions (described in the previous section), and deterministic tractography was performed to identify tracts likely affected by damage to each ROI. The approach is outlined in Figure 5A. Tracts identified for each ROI are shown in Figure 5B along with ROI peak coordinates and statistics.

**4. Discussion**

*4.1 White matter bottleneck damage contributes to chronic language deficits*

Previous studies provided evidence that white matter bottlenecks exist near regions where damage predicts chronic language deficits. However, they did not explicitly show that damage to the bottlenecks in these regions is itself predictive of language deficits. This limits the strengths of inferences that can be drawn regarding the contribution of damage affecting these regions to chronic language deficits. The ROI analyses employed here, by contrast, (1) directly assessed how chronic language deficits relate to the lesioning of white matter bottlenecks in these areas that were identified *a priori*, and (2) demonstrate that damage to these regions is a better predictor of chronic language deficits than both the amount of damage sustained by the tracts that pass through them and the amount of concomitant damage to nearby language areas. These results were further confirmed by our data-driven analyses.

Naming, fluency, and AudSem scores were significantly impaired by damage to the posterior bottleneck region (Figures 3 and 4). This corroborates findings from previous studies that have implicated damage to the white matter near this region in language deficits that span broad domains of language function (Baldo *et al*., 2013; Butler *et al*., 2014; Dronkers *et al*., 2004; Fridriksson, 2010; Fridriksson *et al*., 2013; Geva *et al*., 2012; Harvey and Schnur, 2015; Henseler *et al*., 2014). This region contains fibers associated with both dorsal (AF) and ventral (ILF and IFOF) language processing streams that support integrated sensorimotor language processing and the extraction of meaning from auditory language, respectively (Kümmerer *et al.*, 2013; Saur *et al.*, 2008). Thus, our results support the conclusion that damage to the WM underlying the left pMTG is associated with broad language impairments because it simultaneously disrupts both dorsal and ventral language pathways. Indeed, a recent study reported that a patient with damage confined to the portion of the left posterior temporo-parietal white matter containing fibers associated with these tracts showed profound impairments of speech production despite the absence of frontal damage, providing single-case evidence congruent with our conclusion (Ivanova et al. 2016). Our SVR-LSM+DTLVC analyses, also found that damage to this region is a strong predictor of deficits in all three language outcomes (Figure 4; Table). Congruent with the findings of Turken and Dronkers (2011), our *post-hoc* fiber tracking results suggest that damage to these pSTG/pMTG WM regions likely result in damage to the AF long direct segment, AF posterior indirect segment, ILF, IFOF, and tapetum of the corpus callosum (Figure 5B). While speculative, the incorporation of information regarding lesion status at the posterior bottleneck region could aid the development of future prognostic criteria for patients with left hemispheric stroke.

Our results suggest that damage to the anterior bottleneck predicts deficits in fluency, but do not support the hypothesis that damage to this region leads to deficits in picture naming or auditory comprehension abilities as measured in this study. We note that Mirman and colleagues (2015a, 2015b) found that deficits in semantic recognition, a variable obtained from a factor analysis of scores on a large battery of neuropsychological language measures that featured strong loadings from tests involving verbal and non-verbal semantic categorization and picture naming, were predicted by damage to a white matter cluster partially overlapping with the anterior bottleneck region. While we did not find evidence for a relationship between damage to this region and performance on the BNT (a picture naming test) (Figures 3 and 5), our results suggest that damage to this region may contribute to category fluency deficits (Figures 3 and 4). Notably, the measures employed by Mirman and colleagues (2015a, 2015b) did not include measures analogous to the category fluency measures utilized in this study, and it is possible that the relationship observed by Mirman and colleagues primarily reflects deficits in categorical processing. Indeed, our supplementary analyses suggest that damage to this region predicts deficits in both phonological and semantic category fluency (Supplementary Analysis 1), indicating that deficits following damage to this region are not confined to semantic processes.

Notably, one of the three peak WM lesion predictors of fluency deficits identified by the SVR-LSM+DTLVC analyses partially overlapped with the anterior bottleneck region (Figure 4B). The fiber tracking of the ROI derived from this region provides some clues as to why damage to this region might produce fluency deficits. In addition to fibers associated with the left ATR, UF, and IFOF (i.e. the tracts used to define the anterior bottleneck), fibers associated with the left FAS, cortico-striatal projections, and short-range cortico-cortico fibers connecting left IFGpop and IFGptr also pass through this region (Figure 5B). Previous research indicates that verbal fluency deficits in patients with primary progressive aphasia likely result in part from the degeneration of the FAS, which connects the superior frontal gyrus/pre-supplementary motor area to the IFGpop (Catani *et al.*, 2013). Deficits in both phonological and semantic fluency have been reported following damage to basal ganglia structures that include the caudate and putamen (Biesbroek *et al.*, 2015; Chouiter *et al.*, 2016), and basal ganglia degeneration correlates with fluency deficits in individuals with HIV (Thames *et al.*, 2012). FMRI evidence also implicates the basal ganglia in aspects of language processing that include word generation and speech fluency (Lu *et al.*, 2010; Seghier and Price, 2010; Seghier *et al.*, 2014). Furthermore, inhibitory signaling from the left IFGpop to left IFGptr likely occurs during word generation, and may reflect the application of phonological constraints to lexical retrieval processes during word generation (Heim *et al.*, 2009). Thus, in addition to the ventral pathway and thalamo-cortical fibers passing through the anterior bottleneck region, fibers associated with at least three pathways relevant to verbal fluency may also traverse this region. While these results should be considered tentative until confirmed by future studies, the disruption of processing via these pathways may contribute to verbal fluency deficits following damage to this region.

*4.2 Additional lesion predictors of language deficits*

Our SVR-LSM+DTLVC analyses revealed significant associations between damage to the left pSMG/AG/IFG/ATL and deficits in all three language outcomes, although the precise locations and extents differed for each outcome (Figure 4). Speculatively, this may, in part, reflect the role of these regions (particularly the left pSMG/AG) as heteromodal "information convergence zones" that are important for general semantic processing (Binder and Desai, 2011). Owing to the coarser nature of our language measures relative to those measured by Mirman and colleagues (2015a, 2015b), the ubiquity of these regions as lesion predictors of deficits in the current study may reflect involvement in processes "shared" by all three measures (e.g. associative retrieval, verbal working memory, etc.) (Lau et al., 2015).

Our SVR-LSM+DTLVC findings regarding lesion predictors of naming deficits are highly consistent with those of another recent study that investigated the lesion correlates of overall performance deficits on the same picture naming test employed here (Baldo *et al.*, 2013). This study found significant lesion predictors in anterior, middle, and posterior segments of the left MTG/STG and underlying WM. Other recent studies have also implicated damage to the left anterior temporal lobe, IFG, and pSMG in post-stroke naming deficits (Lau *et al.*, 2015), but damage to the left posterior temporal lobe and underlying white matter are consistently identified as strong lesion predictors of naming deficits (Baldo *et al.*, 2013; Lau *et al.*, 2015; Pustina *et al.*, 2016). Our SVR-LSM+DTLVC findings regarding lesion predictors of fluency and auditory semantic decision-making deficits are also largely consistent with the results of previous studies that have implicated damage to the left IFGpop/PCG, anterior AF, anterior insula, and putamen in post-stroke fluency deficits (Bates *et al.*, 2003; Biesbroek *et al.*, 2015; Chouiter *et al.*, 2016; Fridriksson *et al.*, 2013), and damage to the left IFG, posterior STG/MTG, pSMG/AG, and anterior temporal lobe in deficits of auditory comprehension and/or discrimination (Bates *et al.*, 2003; Dronkers *et al.*, 2004; Fridriksson *et al.*, 2013; Geva *et al.*, 2012; Pustina *et al.*, 2016). Notably, the temporal lobe cortical lesion predictors of deficits in AudSem performance abilities overlapped primarily with pMTG/tocMTG (BA21 -- Figure 4) and overlapped to a lesser degree with the posterior superior temporal sulcus (BA22 – STS), consistent with previous reports that have suggested that the posterior middle temporal cortex may be more important for auditory semantic comprehension than canonical Wernicke's area (Dronkers *et al*., 2004). Thus, our SVR-LSM+DTLVC results are largely consistent with the findings of previous lesion-deficit studies, and provide additional confirmation for the results of our ROI analyses.

*4.3 Spatially specific white matter damage and controls for concomitant damage*

      The results from our ROI analyses indicate that the observed relationships between chronic language deficits and lesions affecting white matter bottlenecks are not driven by the amount of damage sustained by the tracts that pass through them or by damage to nearby cortical language areas (Figure 3). As previously noted, few studies investigating relationships between tract damage and language outcome have attempted to control for concomitant damage to other tracts or for cortical damage, weakening the conclusions that can be drawn regarding the effects of damage to the tracts being investigated. As our findings indicate that the effects of damage to long-range white matter pathways on language outcomes may be spatially specific, it is worth considering that spatially non-discriminative measures such as tract lesion load may not be ideal for characterizing lesion-symptom relationships. Indeed, a similar argument regarding the importance of developing more informative characterizations of tract damage for lesion-behavior analyses was made by recent study comparing measures of lesion load and expected tract disconnection as predictors of chronic deficits in naming and fluency (Hope *et al*., 2015). In the current study, despite the considerable explanatory power of some of the control models (Figure 3A), lesion status information for the relevant white matter bottleneck ROIs significantly improved all models as indicated by the $R^2$ change analysis (Figure 3). This highlights both the detrimental nature of damage to these regions and the shortcomings of spatially non-specific measures of tract damage, and supports the application of more spatially discriminative measures of tract damage.

*4.4 Limitations and conclusions*

      A limitation of the current study is that the bottleneck ROIs were defined using thresholded tract probability maps. Thus, it might be argued that the results of our ROI analyses may differ depending on threshold choice. However, the fact that similar results

were obtained from a data-driven analysis (SVR-LSM) and *post-hoc* fiber tracking indicates that our results are robust to changes in analytical strategy. This is also supported by our supplemental analyses employing alternate ROI definitions (Supplementary Material 4). A second limitation is that only chronic patients were included, and this prevents drawing strong conclusions about how damage to the bottleneck regions affects the success of long-term recovery vs. chronic aphasia severity, per se (i.e. it is unclear if damage to these regions causes severe long-term deficits, whether it impedes recovery, or both). Studies in patients presenting with acute symptoms and longitudinal studies of recovery are necessary allow for such conclusions. Lastly, it is not clear how damage to these regions affects language network function, although it would be expected that associated disruptions of connectivity, particularly by damage to the posterior temporal region, might impede typical function in left hemispheric language networks and lead to the recruitment of atypical networks to accomplish language processing. This should be addressed by future studies.

Our data support the conclusion that WM bottlenecks correspond to structural weak points in the neural architecture of the distributed language network. Lesions affecting the posterior bottleneck are associated with poor long-term prognosis and lead to chronic deficits in both expressive and receptive language. Lesions affecting the anterior bottleneck are primarily associated with chronic deficits in expressive language functions, although future work is needed to fully understand the specific effects of damage to this region. Our results emphasize the importance of considering the effects of spatially specific white matter damage in patients with aphasia following left hemisphere stroke.

**Acknowledgements**
Amber Martin
Christi Banks
Michel Thiebaut de Schotten for helpful discussion and for providing the atlas labels used for alternate ROI definitions in the supplementary analyses.
Data were provided [in part] by the Human Connectome Project, WU-Minn Consortium (Principal Investigators: David Van Essen and Kamil Ugurbil; 1U54MH091657) funded by the 16 NIH Institutes and Centers that support the NIH Blueprint for Neuroscience Research; and by the McDonnell Center for Systems Neuroscience at Washington University.

**Funding**

NIH R01 HD068488

NIH R01 NS048281

**Tables**

**Table. SVR-LSM+DTLVC peak and cluster statistics**

| Peak Location | Extent | beta | x | y | z |
|---|---|---|---|---|---|
| **Cluster Peaks for SVR-LSM+DTLVC of Fluency Deficits** | | | | | |
| Posterior Middle Temporal WM | 3869 | -0.87 | -30 | -48 | 14 |
| Precentral WM | 3869 | -0.68 | -42 | -6 | 28 |
| Posterior Supramarginal Gyrus | 3869 | -0.58 | -60 | -46 | 26 |
| Callosal WM | 22 | -0.48 | -22 | -12 | 28 |
| Postcentral Gyrus | 179 | -0.44 | -64 | -12 | 16 |
| Anterior Middle Temporal Gyrus | 179 | -0.37 | -60 | -8 | -16 |
| Callosal WM | 15 | -0.40 | -22 | -22 | 34 |
| Precuneate WM | 11 | -0.31 | -24 | -48 | 42 |
| **Cluster Peaks for SVR-LSM+DTLVC of Naming Deficits** | | | | | |
| Posterior Middle Temporal WM | 536 | -1.04 | -38 | -40 | 2 |
| Posterior Supramarginal WM | 536 | -0.79 | -40 | -46 | 32 |
| Precentral WM | 25 | -0.73 | -42 | -6 | 28 |
| Precentral Gyrus | 175 | -0.71 | -40 | 6 | 26 |
| Postcentral Gyrus | 38 | -0.71 | -64 | -12 | 16 |
| Planum Polare | 72 | -0.69 | -42 | -16 | -4 |
| Posterior Supramarginal Gyrus | 164 | -0.66 | -58 | -48 | 22 |
| Anterior Supramarginal Gyrus | 164 | -0.41 | -64 | -24 | 40 |
| Parietal Operculum | 32 | -0.66 | -46 | -30 | 24 |
| Anterior Middle Temporal WM | 123 | -0.65 | -44 | -2 | -24 |
| Inferior frontal gyrus PTr WM | 15 | -0.62 | -38 | 34 | 4 |
| Anterior Middle Temporal Gyrus | 80 | -0.55 | -60 | -8 | -16 |
| Anterior Superior Temporal Gyrus | 10 | -0.45 | -62 | -6 | -2 |
| **Cluster Peaks for SVR-LSM+DTLVC of Auditory Semantic Decision Deficits** | | | | | |
| Posterior Middle Temporal WM | 1475 | -1.10 | -34 | -46 | 8 |
| Posterior Supramarginal Gyrus | 1475 | -0.90 | -58 | -46 | 28 |
| Inferior Frontal Gyrus PTr | 11 | -0.68 | -52 | 26 | 18 |
| Anterior Middle Temporal Gyrus | 33 | -0.59 | -60 | -8 | -16 |

**Figures**

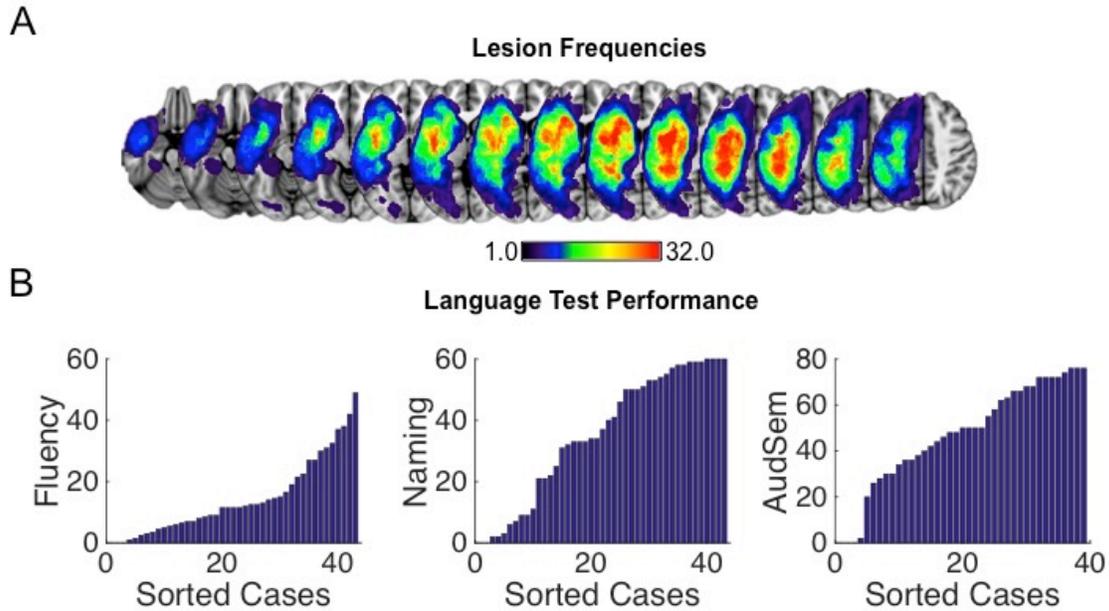

**Figure 1. Characterization of lesion and language test data**. **A.** Lesion frequency overlaps are shown on axial slices from a template brain. Colorbar values range from 1 to 32, and indicate the number of patients with lesions at each voxel. **B.** Sorted scores (from low to high) are shown for the Fluency (left), Naming (middle), and Auditory Semantic Decision (AudSem -- right) tasks.

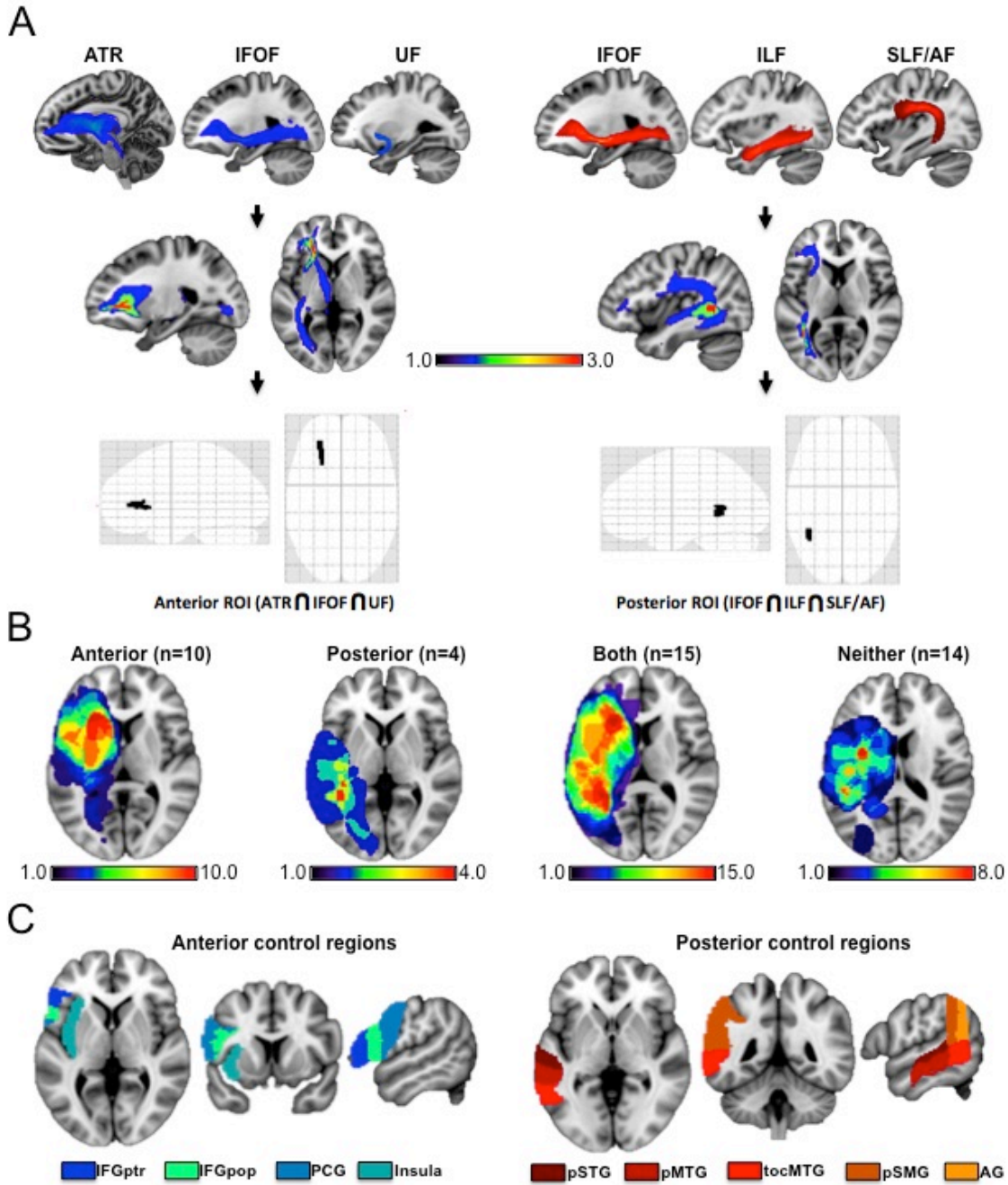

**Figure 2. Regions of interest. A.** The process for defining each ROI is illustrated in three steps. First, a threshold of 20% was applied to the atlas labels corresponding to the left hemispheric tracts expected to pass through the anterior (top -- blue) and posterior (top -- red) bottleneck ROIs. Next, the thresholded tracts were then binzarized and overlapped (middle – colorbar values indicate number of tracts at each voxel). Finally, the anterior bottleneck ROI was defined as the intersection of the voxels contained in the thresholded ATR, IFOF, and UF labels (bottom left), and the posterior bottleneck ROI was defined as the intersection of the voxels contained in the thresholded IFOF, ILF, and SLF/AF labels (bottom right). **B.** Representative slices overlaid with lesion-frequency maps for patients with damage to only the anterior bottleneck ROI (left), only the posterior bottleneck ROI (left middle), both bottleneck ROIs (right middle), and neither bottleneck ROI (right) are shown. **C.** Harvard-Oxford cortical atlas labels for the left inferior frontal gyrus pars triangularis (IFGptr) and pars opercularis (IFGpop), precentral gyrus (PCG), and insula (left) were used to define the cortical control regions for the anterior bottleneck ROI (left). Labels for the left posterior superior and middle temporal gyri (pSTG/pMTG), temporo-occipital middle temporal gyrus (tocMTG), posterior supramarginal gyrus (pSMG), and angular gyrus (AG) were used to define the cortical control regions for the posterior bottleneck ROI (right). Cortical labels were thresholded to contain only voxels with grey matter tissue probabilities greater than 20%. *Note:* ∩ *indicates the intersection operation*

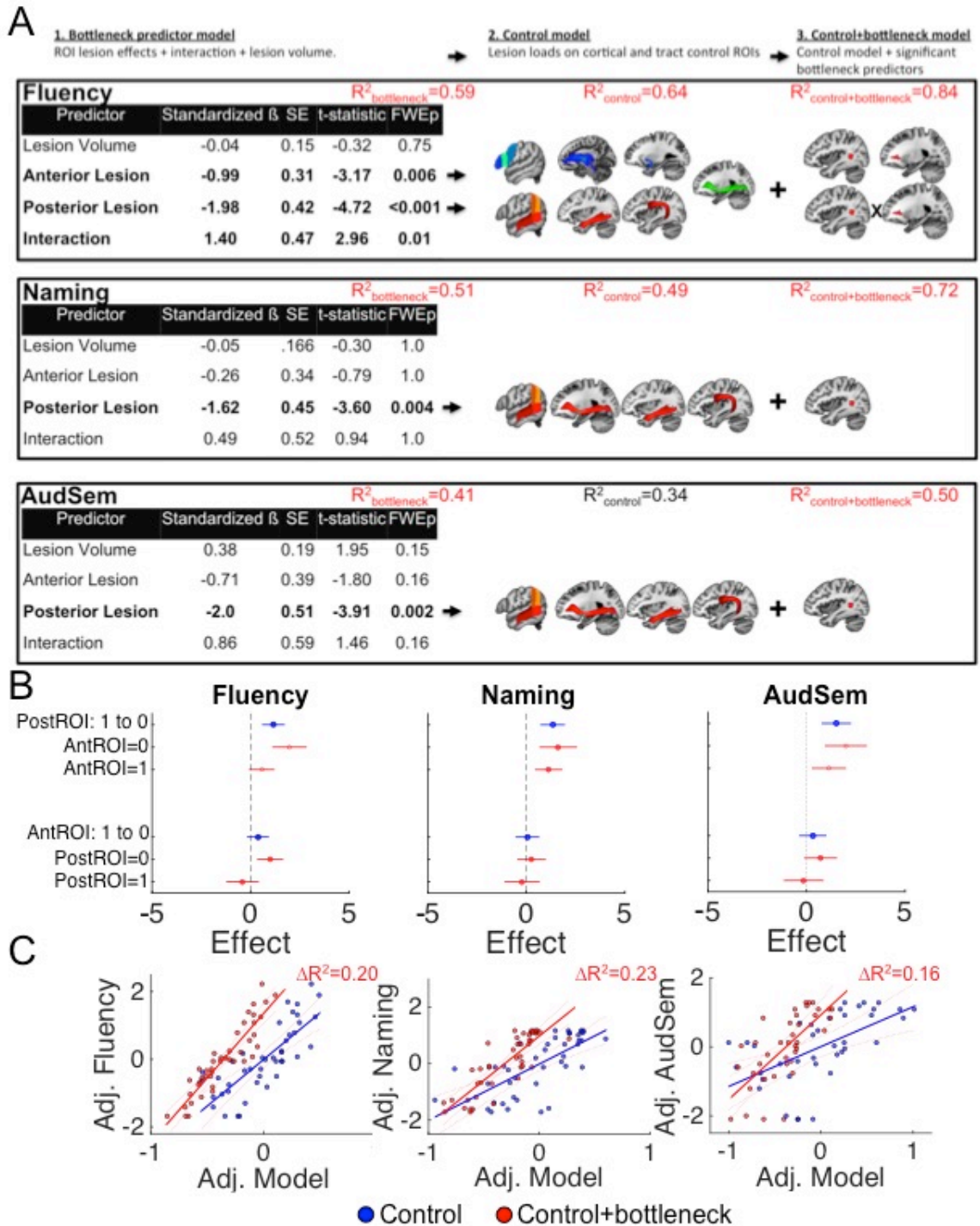

**Figure 3. Multiple regression results. A.** For each language outcome, tables containing statistics for predictors from the bottleneck predictor model are shown (left – note: bold table text indicates p<0.05, FWE-corrected), along with graphical depictions illustrating the variables entered in the first blocks (control model -- middle) and second blocks

(control+bottleneck model -- right) of the hierarchical regressions performed to control for cortical and tract damage effects. The $R^2$ for each model is also shown (note: red $R^2$ text indicates p<0.05, FWE-corrected). **B.** Bottleneck ROI effects from the bottleneck predictor models are shown. Blue data points illustrate the estimated effects (and 95% CIs) of changing lesion status at each bottleneck ROI from lesioned (1) to un-lesioned (0) on each language outcome, after averaging out the effects of other predictors. Red data points illustrate the estimated effects (and 95% CIs) for each bottleneck ROI when the other bottleneck ROI is lesioned (1) or unlesioned (0). Note: parameter estimates for each bottleneck ROI shown in (A) correspond to the effects of damage when the other ROI is not lesioned (i.e. = 0). **C.** For each language outcome, regression fits for the control model (shown in blue) and control+bottleneck model (shown in red) are shown along with the change in $R^2$ between the two models (note: red $R^2$ text indicates p<0.05, FWE-corrected).

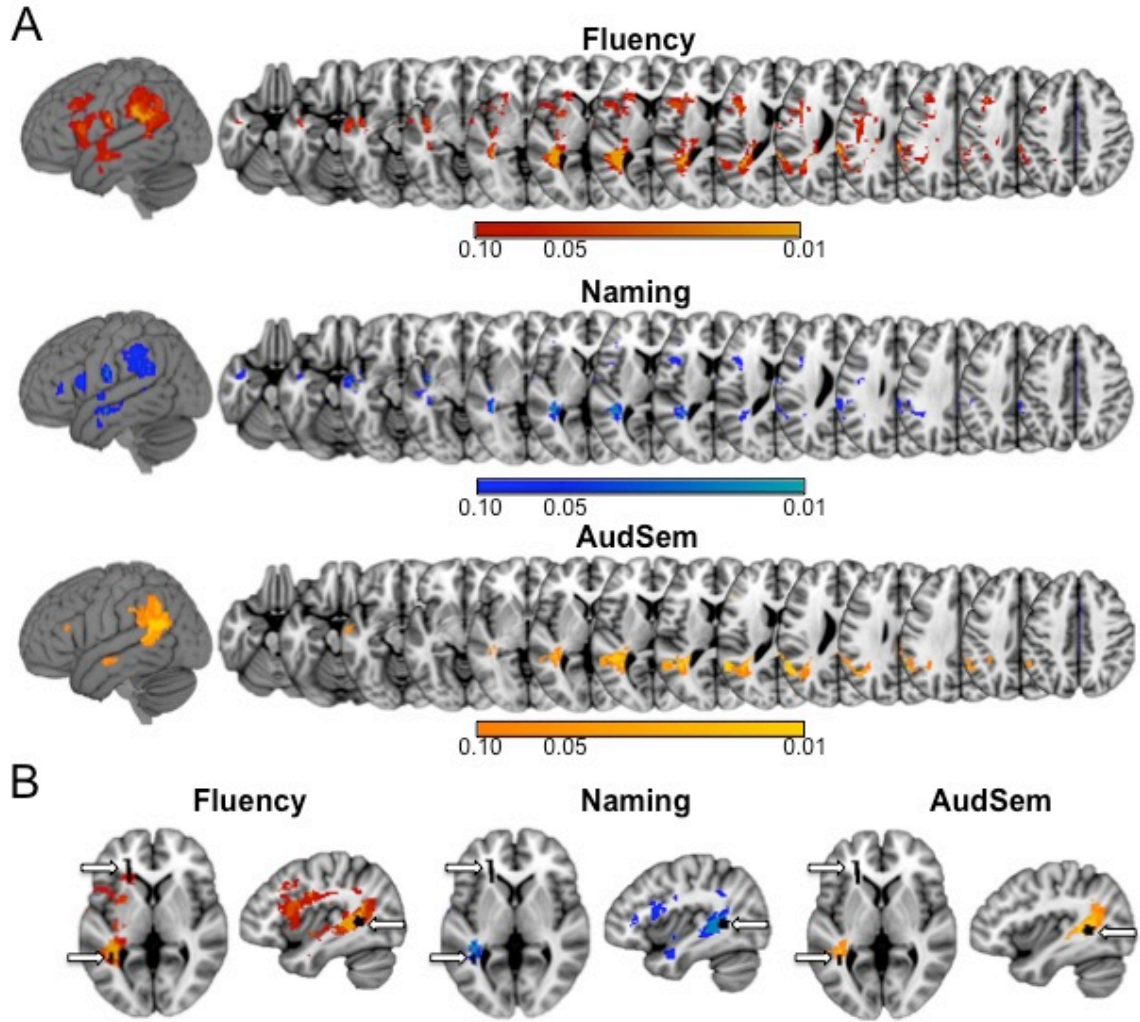

**Figure 4. SVR-LSM+DTLVC Results. A.** Left hemisphere lesion predictors of deficits in fluency (top), naming (middle), and auditory semantic decision (bottom) abilities. Each map shows FDR-corrected p-values (ranging from 0.1 to 0.01) obtained from permutation testing of the SVR-LSM+DTLVC model (2000 permutations). B. Overlap of left hemispheric lesion predictors of language deficits identified by SVR-LSM+DTLVC and white matter bottleneck ROIs. White matter bottleneck ROIs are shown in black and indicated by arrows.

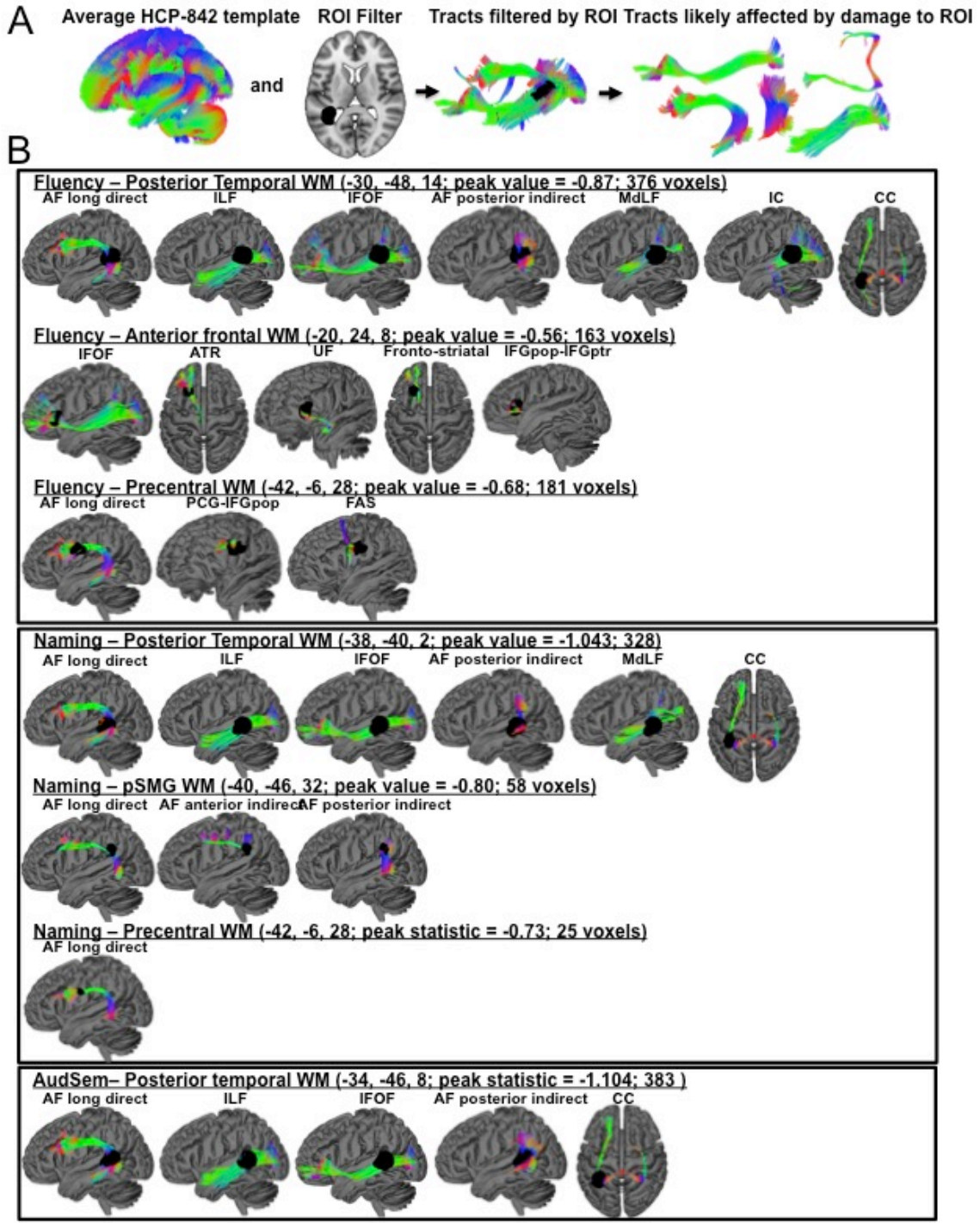

**Figure 5. Deterministic tractography results. A.** A schematic diagram illustrating the deterministic tractography approach is shown. Fibers from the average HCP-842 template with termination points in the grey matter were filtered to leave only fibers that passed through each left hemisphere ROI. The resulting set of fibers was then segmented into constituent tracts that were labeled according to previous reports. **B.** 3D brain renderings show each peak white matter ROI (black regions) and the fiber tracts identified by the deterministic tractography analyses. Tracts associated with three ROIs are shown for Fluency (top panel) and Naming (middle panel), and tracts associated with one ROI are shown for AudSem (bottom panel).

**Supplementary Material**

**Supplementary Table 1**

Supplementary Table 1. Patient characteristics

| Patient | Age | Sex | EHI | TSS | BNT | COWAT | SFT | Average Fluency | %SD Correct |
|---|---|---|---|---|---|---|---|---|---|
| 1 | 63 | F | 0.55 | 1.0 | 59 | 27 | 47 | 37 | 68 |
| 2 | 78 | F | 1.00 | 4.1 | 57 | 21 | 39 | 30 | 50 |
| 3 | 41 | F | 0.50 | 5.8 | 9 | 4 | 9 | 6.5 | 38 |
| 4 | 54 | M | 1.00 | 1.6 | 7 | 5 | 6 | 5.5 | 66 |
| 5 | 46 | M | 0.90 | 1.0 | 53 | 20 | 42 | 31 | 46 |
| 6 | 52 | M | 0.58 | 1.0 | 60 | 11 | 27 | 19 | 48 |
| 7 | 56 | M | 1.00 | 3.4 | 32 | 2 | 14 | 8 | 50 |
| 8 | 53 | M | 1.00 | 5.0 | 50 | 8 | 20 | 14 | 76 |
| 9 | 55 | M | 1.00 | 1.2 | 58 | 15 | 30 | 22.5 | 72 |
| 10 | 48 | M | 1.00 | 6.1 | 22 | 0 | 12 | 6 | 20 |
| 11 | 63 | M | 1.00 | 1.0 | 60 | 8 | 10 | 9 | 40 |
| 12 | 56 | F | 1.00 | 1.0 | 33 | 6 | 20 | 13 | NA |
| 13 | 23 | M | 1.00 | 1.0 | 60 | 36 | 62 | 49 | 76 |
| 14 | 50 | M | 1.00 | 1.0 | 2 | 0 | 2 | 1 | 28 |
| 15 | 48 | F | 1.00 | 1.0 | 60 | 24 | 60 | 42 | 72 |
| 16 | 70 | F | 1.00 | 2.0 | 11 | 3 | 3 | 3 | 0 |
| 17 | 68 | M | 0.91 | 3.3 | 9 | 4 | 10 | 7 | 50 |
| 18 | 59 | M | 0.82 | 1.0 | 53 | 19 | 35 | 27 | 72 |
| 19 | 23 | F | 1.00 | 1.0 | 59 | 20 | 45 | 32.5 | 76 |
| 20 | 24 | F | 1.00 | 1.0 | 59 | 31 | 45 | 38 | 58 |
| 21 | 78 | F | 1.00 | 3.4 | 58 | 9 | 21 | 15 | 36 |
| 22 | 65 | M | 1.00 | 14.0 | 55 | 18 | 36 | 27 | 68 |
| 23 | 58 | F | 1.00 | 13.0 | 40 | 13 | 12 | 12.5 | 36 |
| 24 | 72 | F | 1.00 | 1.5 | 0 | 0 | 0 | 0 | NA |
| 25 | 50 | M | 1.00 | 2.9 | 0 | 0 | 0 | 0 | 42 |
| 26 | 57 | M | 1.00 | 2.1 | 2 | 2 | 1 | 1.5 | 48 |
| 27 | 51 | M | 1.00 | 1.1 | 37 | 8 | 15 | 11.5 | 66 |
| 28 | 43 | M | 1.00 | 1.3 | 50 | 11 | 22 | 16.5 | 44 |

| | | | | | | | | |
|---|---|---|---|---|---|---|---|---|
| 29 | 24 | M | 0.83 | 2.3 | 21 | 9 | 14 | 11.5 | 2 |
| 30 | 67 | F | 1.00 | 2.2 | 6 | 2 | 3 | 2.5 | 50 |
| 31 | 62 | F | -1.00 | 4.4 | 33 | 20 | 23 | 21.5 | 62 |
| 32 | 44 | F | 0.91 | 2.1 | 41 | 7 | 18 | 12.5 | 30 |
| 33 | 62 | M | 1.00 | 2.6 | 54 | 10 | 19 | 14.5 | 63 |
| 34 | 31 | M | 1.00 | 4.8 | 21 | 7 | 7 | 7 | NA |
| 35 | 61 | M | 1.00 | 9.6 | 25 | 1 | 6 | 3.5 | NA |
| 36 | 64 | M | -1.00 | 2.7 | 51 | 1 | 16 | 8.5 | 30 |
| 37 | 38 | F | 0.91 | 1.8 | 46 | 11 | 12 | 11.5 | 72 |
| 38 | 53 | F | 1.00 | 9.2 | 34 | 2 | 7 | 4.5 | 74 |
| 39 | 54 | M | 0.92 | 3.3 | 33 | 5 | 19 | 12 | 55 |
| 40 | 46 | M | 1.00 | 1.3 | 31 | 1 | 9 | 5 | 34 |
| 41 | 90 | F | 0.71 | 1.3 | 3 | 0 | 0 | 0 | 0 |
| 42 | 29 | F | 1.00 | 3.4 | 50 | 4 | 19 | 11.5 | 0 |
| 43 | 67 | M | 1.00 | 12.4 | 34 | 2 | 16 | 9 | 26 |

*EHI – Edinburgh Handedness Inventory, TSS – time since stroke, BNT – Boston Naming Test, COWAT – Controlled Oral Word Association Test, SFT – Semantic Fluency Test, %SD Correct -- % Semantic Decision Correct.
*Note:* The average fluency score is the average of the COWAT and SFT.

**Supplementary Analysis 1**

The multiple regression analyses were performed using the COWAT and SFT (i.e. measures of phonological and semantic fluency, respectively) as separate outcomes, rather than the average fluency measure reported in the main text. Model 1 significantly predicted deficits on the COWAT ($R^2=0.58$, $F_{4,38}=13.2$, $p<0.001$). Posterior ROI lesion status ($\beta=-1.8$, $t=-4.23$, $p<0.001$), anterior ROI lesion status ($\beta=-1.09$, $t=-3.46$, $p=0.001$), and the interaction term ($\beta=1.41$, $t=2.95$, $p=0.005$) uniquely predicted deficits on the COWAT. Model 2a also significantly predicted deficits on the COWAT ($R^2=0.61$, $F_{14,28}=3.19$, $p=0.004$). For Model 2b, the addition of significant terms from Model 1 significantly improved model fit ($\Delta R^2=0.12$, $F_{3,25}=3.70$, $p=0.02$).

Model 1 significantly predicted deficits on the SFT ($R^2=0.54$, $F_{4,38}=11.2$, $p<0.001$). Posterior ROI lesion status ($\beta=-1.72$, $t=-3.91$, $p<0.001$), anterior ROI lesion status ($\beta=-1.01$, $t=-3.07$, $p=0.003$), and the interaction term ($\beta=1.35$, $t=2.72$, $p=0.009$) uniquely predicted deficits on the COWAT. Model 2a also significantly predicted deficits on the COWAT ($R^2=0.59$, $F_{14,28}=2.99$, $p=0.007$). For Model 2b, the addition of significant terms from Model 1 significantly improved model fit ($\Delta R^2=0.20$, $F_{3,25}=7.37$, $p<0.001$).

Thus, nearly identical conclusions are supported by separate ROI analyses of each fluency measure.

**Supplementary Analysis 2**

The SVR-LSM+DTLVC analyses were performed using the COWAT and SFT rather than the average fluency measure reported in the main text. Results are shown in Supplementary Figure 1.

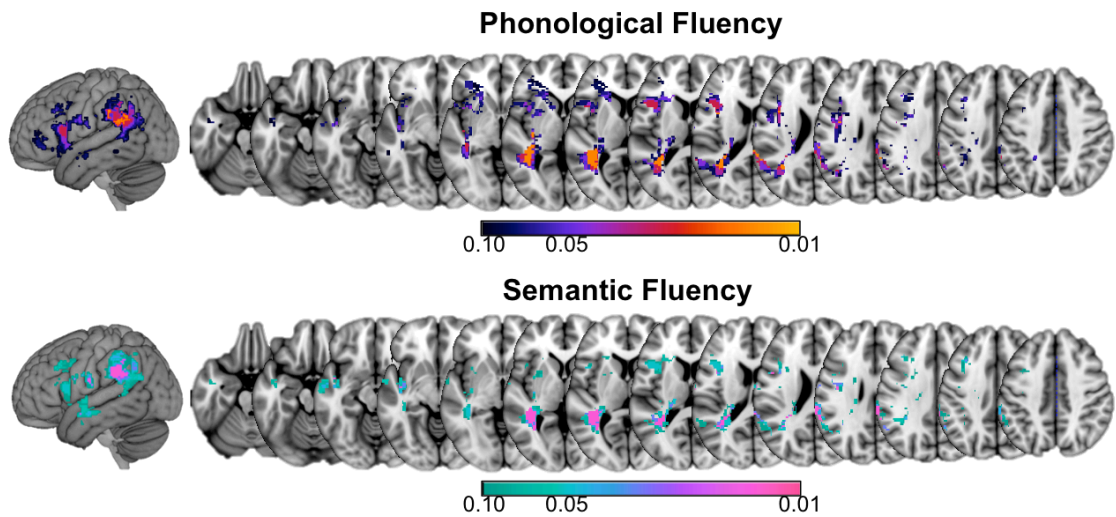

**Supplementary Figure 1. SVR-LSM+DTLVC Results for the SFT and COWAT.** Significant (FDR<0.10) lesion predictors of performance on the COWAT (Phonological fluency -- top) and SFT (Semantic fluency -- bottom) are shown. Note that the overall patterns are highly similar to each other and to the results for the average fluency measure reported in the main text. Notably, damage to the anterior bottleneck region appears more pronounced for the COWAT (top), whereas damage to the ATL appears more pronounced for the SFT (Bottom).

**Supplementary Analysis 3**

The SVR-LSM analyses were performed without DTLVC. Results are shown in Supplementary Figure 2.

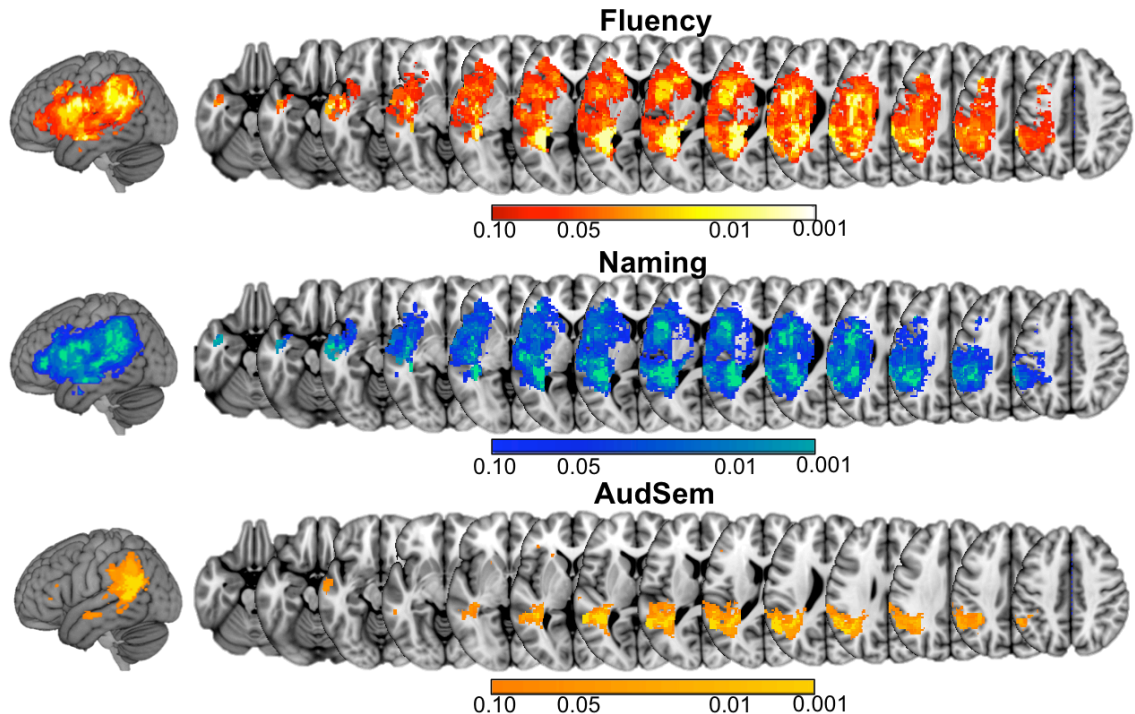

**Supplementary Figure 2. SVR-LSM results for Fluency, Naming, and AudSem without DTLVC.** Significant (FDR<0.10) lesion predictors of performance on the average fluency measure (top), picture naming (middle), and auditory semantic decisions (bottom) identified by the SVR-LSM analyses without DTLVC are shown. Note that while the results appear less specific (due to increased power and a higher number of significant voxels), the most significant regions for each outcome correspond closely to the regions identified by the analyses with DTLVC reported in the main text. Colorbar values indicate FDR-corrected p-values, and range from the threshold limit (0.10) to 0.001 (the most reliable effects in each map).

**Supplementary Analysis 4**

To assess whether similar results might be obtained using ROIs defined using a different atlas with different thresholds, we defined a second set of bottleneck ROIs using a recently published probabilistic tractography atlas (Rojkova et al., 2016 – available at brainconnectivitybehavior.eu). Because these labels were much less sparse than those used for the primary analysis and extended to the pial surface at thresholds used in the primary analyses (i.e. see Supplementary Figure 3A), a more conservative threshold of 70% was applied. The ROIs were defined using labels corresponding to the same tracts described in the main report, with the exception that a label corresponding to the AF long direct segment was used as opposed to the combined SLF/AF label employed in the main report. The thresholded labels and resulting ROIs are shown in Supplementary Figure 3B/C. The posterior ROI defined using the Rojkova atlas partially overlapped with the one defined using the ICBM-DTI atlas, but was situated slightly more laterally and closer to the superficial white matter (Supplementary Figure 3C). The anterior ROI defined using the Rojkova atlas partially overlapped with the one defined using the ICBM-DTI atlas, but was situated slightly more ventrally and extended further in both anterior-posterior and medial-lateral planes (Supplementary Figure 3C).

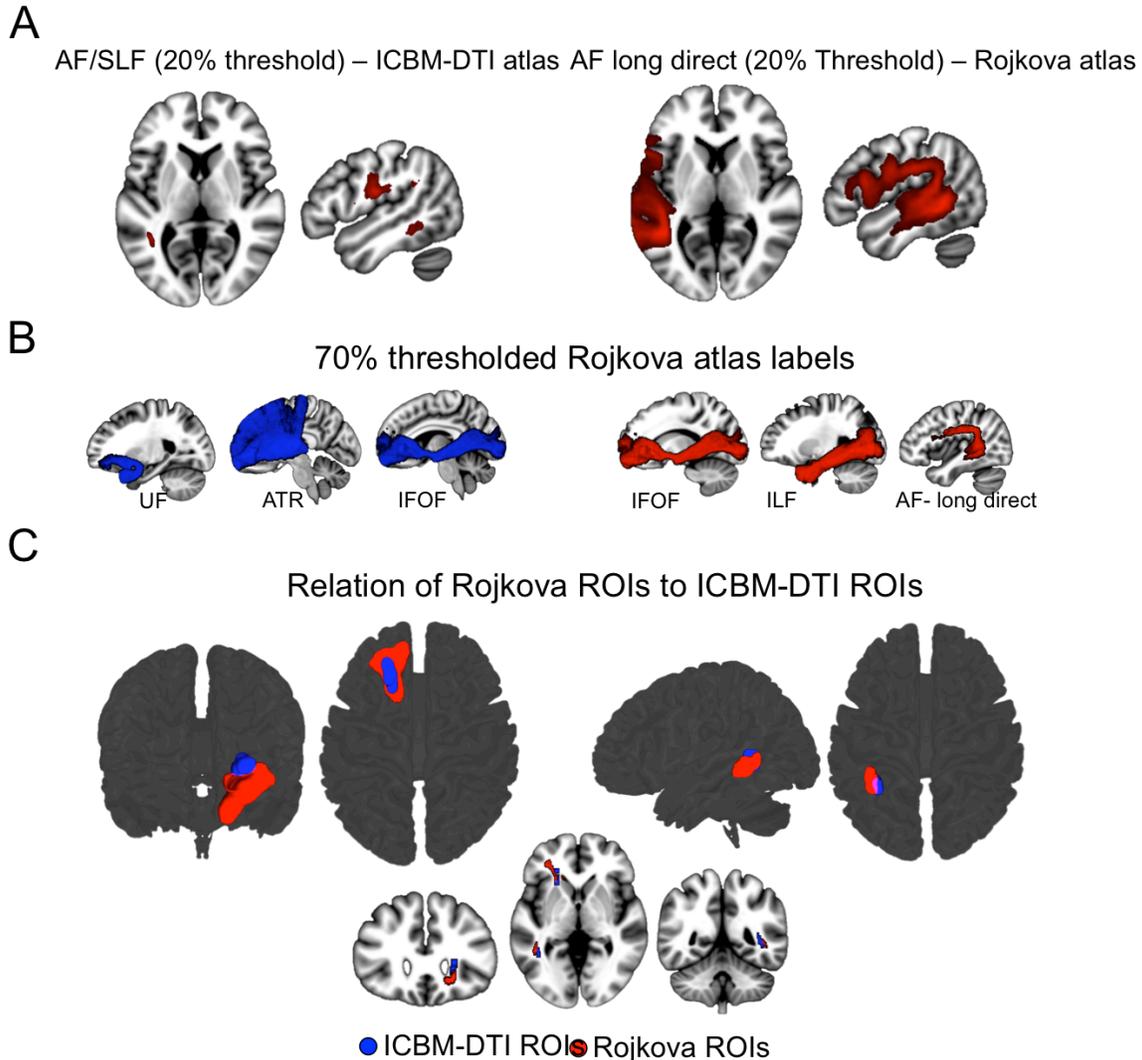

**Supplementary Figure 3**. **Alternate ROI definitions using the atlas described by Rojkova et al (2016), and relationships to the ROI definitions used in the primary analyses. A.** 20% thresholed AF labels from the ICBM-DTI atlas used in the main analyses (left) and the Rojkova atlas (right). Note that the labels provided in the Rojkova atlas extend to the pial surface when a 20% probability threshold is employed. B. 70% thresholded labels from the Rojkova atlas were used to define alternate bottleneck ROIs. C. 3D brain renderings and single slice renderings illustrating the similarities and differences between the ROIs defined using the ICBM-DTI atlas (i.e. those used in the initial analyses) and those defined using the Rajkova atlas (i.e. those used in the supplemental analysis). Note that the anterior ROI defined using the Rojkova atlas is much larger than the one defined using the ICBM-DTI atlas, and that the posterior ROI defined using the Rojkova atlas is slightly larger, more lateral, an more anterior than the one defined using the ICBM-DTI atlas.

Hierarchical multiple regression analyses were performed as described in the main manuscript, but using the ROIs defined with the Rojkova atlas rather than the ICBM-DTI atlas. Tract lesion load predictors were also defined using the Rojkova atlas. The purpose of this supplementary analyses was simply to confirm that even when alternate ROI definitions are employed, damage to the bottleneck regions identified in the main analyses provide unique information about language deficits when tract and cortical damage are accounted for.

The bottleneck predictor model for fluency predicted deficits ($R^2=0.50$, $F_{4,38}=9.58$, $p<0.001$). The control model (containing tract and cortical lesion load predictors for the posterior and anterior bottleneck ROIs, since these were identified as predictors of fluency deficits in the primary analysis – tract loads were calculated using the Rojkova atlas) also predicted deficits ($R^2=0.51$, $F_{14,28}=2.11$, $p=0.04$).
The addition of the bottleneck predictors (posterior status, anterior status, and interaction term, since these were identified in the main analysis) improved model fit ($R^2=0.73$, $\Delta R^2=0.22$, $F_{3,25}=7.05$, $p=0.001$).

The bottleneck predictor model for naming predicted deficits ($R^2=0.44$, $F_{4,38}=7.51$, $p<0.001$). The control model (containing tract and cortical lesion load predictors for the posterior bottleneck ROI, since it was identified as significant in the main analysis– tract loads were calculated using the Rojkova atlas) also predicted naming deficits ($R^2=0.46$, $F_{8,34}=3.63$, $p=0.003$). The addition of the posterior bottleneck predictor (since this was identified in the main analysis) improved model fit ($R^2=0.66$, $\Delta R^2=0.20$, $F_{1,33}=19.4$, $p<0.001$).

The bottleneck predictor model for AudSem predicted deficits ($R^2=0.50$, $F_{4,34}=8.42$, $p<0.001$). The control model (containing tract and cortical lesion load predictors for the posterior bottleneck ROI, since it was identified as significant in the main analysis– tract loads were calculated using the Rojkova atlas) did not reliably predict AudSem deficits ($R^2=0.32$, $F_{8,30}=1.75$, $p=0.13$). The addition of the posterior bottleneck predictor (since this was identified in the main analysis) improved model fit ($R^2=0.54$, $\Delta R^2=0.22$, $F_{1,29}=13.87$, $p<0.001$).

Thus, even when an alternate ROI definition is employed, the results are highly similar to those obtained from the original analysis. Critically, the alternately defined bottleneck ROI predictors corresponding to the significant predictors identified by the main analyses still provide unique information about language deficits that are not attributable to damage to the tracts that pass through them or to lesion loads on nearby language-relevant cortices.